\documentclass[journal=nalefd,manuscript=article]{achemso}
\usepackage{chemformula} 
\usepackage[T1]{fontenc} 
\usepackage{amsfonts}
\usepackage{bm}
\usepackage{multirow}
\usepackage{color}
\usepackage{multirow}
\usepackage{epsfig}
\usepackage{graphicx}
\usepackage[normalem]{ulem}
\usepackage{amsmath}
\usepackage{amssymb}
\usepackage{bm}
\usepackage{dcolumn}
\usepackage{braket}
\usepackage{bbold}
\usepackage{natbib}

\usepackage{array}
\usepackage{tabulary}
\usepackage{multirow}
\newcolumntype{C}[1]{>{\centering\arraybackslash}p{#1}}
\newcolumntype{L}[1]{>{\flushleft\arraybackslash}p{#1}}

\author{Jiaqi Feng}
\affiliation{Laboratory of Quantum Functional Materials Design and Application, School of Physics and Electronic Engineering, Jiangsu Normal University, Xuzhou 221116, China}

\author{Xiaodong Zhou}
\email{xdzhou323@gmail.com}
\affiliation{Laboratory of Quantum Functional Materials Design and Application, School of Physics and Electronic Engineering, Jiangsu Normal University, Xuzhou 221116, China}

\author{Meiling Xu}
\affiliation{Laboratory of Quantum Functional Materials Design and Application, School of Physics and Electronic Engineering, Jiangsu Normal University, Xuzhou 221116, China}

\author{Jingming Shi}
\affiliation{Laboratory of Quantum Functional Materials Design and Application, School of Physics and Electronic Engineering, Jiangsu Normal University, Xuzhou 221116, China}

\author{Yinwei Li}
\email{yinwei_li@jsnu.edu.cn}
\affiliation{Laboratory of Quantum Functional Materials Design and Application, School of Physics and Electronic Engineering, Jiangsu Normal University, Xuzhou 221116, China}

\title{Layer Control of Magneto-Optical Effects and Their Quantization in Spin-Valley Splitting  Antiferromagnets}


\begin{document}

	\begin{tocentry}
	
	\includegraphics[width=8.25 cm,height=4.45 cm]{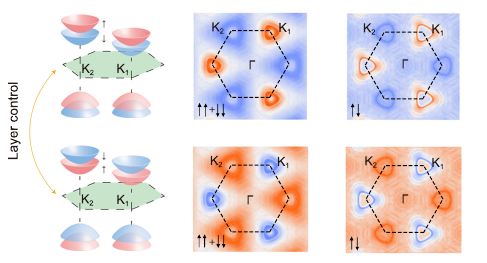}
	
	%
	%
	%
	%
	\end{tocentry}
	
	\newpage

\begin{abstract}
Magneto-optical effects (MOE), interfacing the fundamental interplay between magnetism and light, have served as a powerful probe for magnetic order, band topology, and valley index. Here, based on multiferroic and topological bilayer antiferromagnets (AFMs), we propose a layer control of MOE (L-MOE), which is created and annihilated by layer-stacking or an electric field effect. The key character of L-MOE is the sign-reversible response controlled by ferroelectric polarization, the N{\'e}el vector, or the electric field direction. Moreover, the sign-reversible L-MOE can be quantized in topologically insulating AFMs. We reveal that the switchable L-MOE originates from the combined contributions of spin-conserving and spin-flip interband transitions in spin-valley splitting AFMs, a phenomenon not observed in conventional AFMs. Our findings bridge the ancient MOE to the emergent realms of layertronics, valleytronics, and multiferroics, and may hold immense potential in these fields.
\end{abstract}

	\newpage

\begin{figure}
\centering
	\includegraphics[width=0.65\columnwidth]{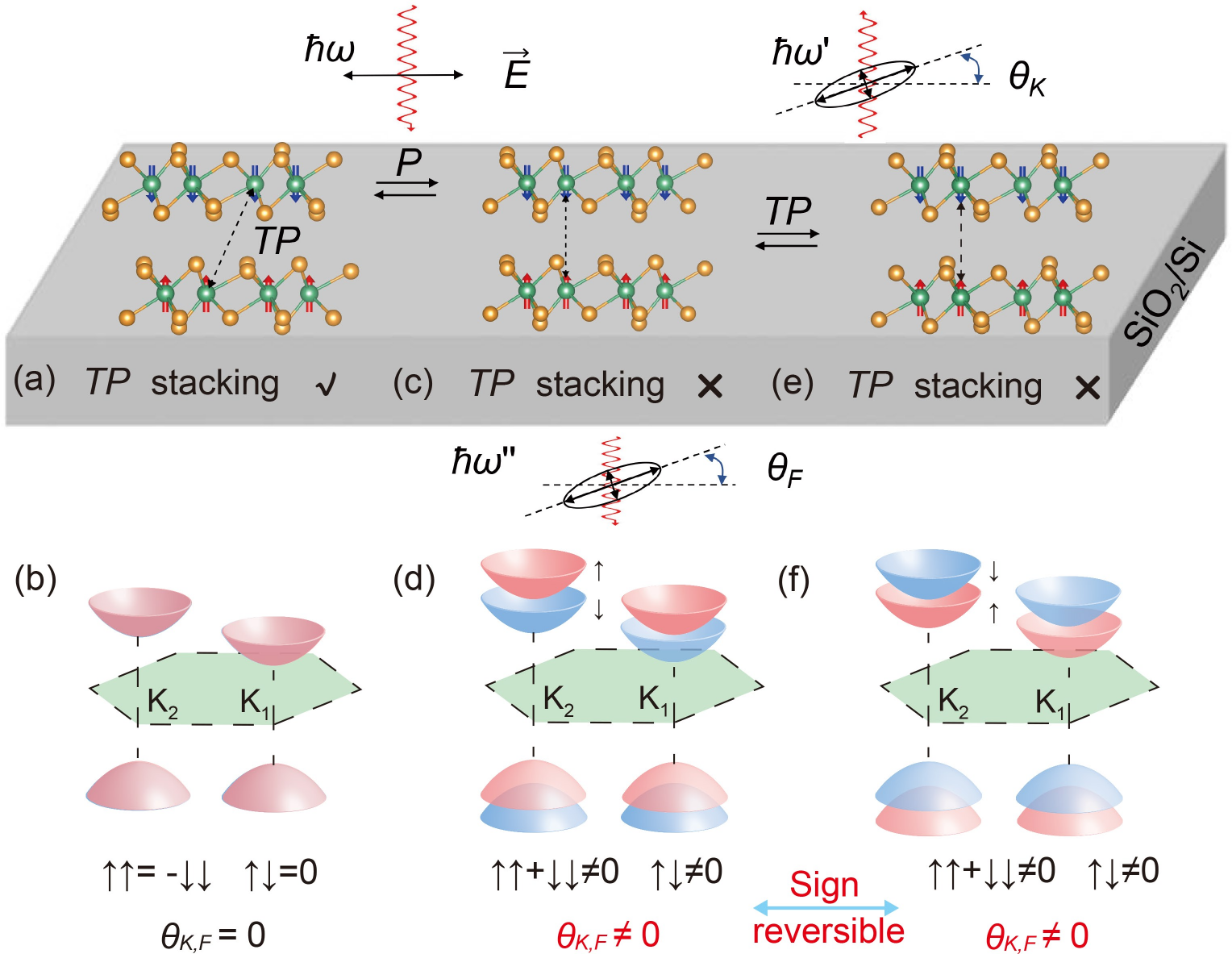}
	\caption{Schematic illustration of L-MOE in spin-valley splitting AFMs. (a-b) The vanishing MOE with conventional $\mathcal{TP}$-symmetric layer stacking. The system exhibits a valley-polarized property, while it still obeys the Kramers spin degeneracy theorem. In this case, the interband optical transition between the spin-up valley and the spin-up valley ($\uparrow\uparrow$) tends to cancel out that of the time-reversal symmetric transition from the spin-down valley to the spin-down valley ($\downarrow\downarrow$). Moreover, the transition between states with opposite spins (spin-flip, $\uparrow\downarrow$) is forbidden. (c-d) After breaking the $\mathcal{TP}$ symmetry via a layer-stacking effect [$e.g.$, from (a) to (c) by rearranging the top layer under a $\mathcal{P}$ operation], the spin degeneracy is further lifted. The combined spin-valley polarization permits the optical excitation from both the spin-conserving ($\uparrow\uparrow$+$\downarrow\downarrow$ $\neq 0$) and spin-flip transitions  ($\uparrow\downarrow$ $\neq 0$). Interestingly, the sign of the L-MOE is reversible by flipping the spin-splitting bands [$e.g.$, from (d) to (f)], which can be achieved by reversing the ferroelectric polarization or N{\'e}el vector direction [$e.g.$, from (c) to (e) by rearranging the bilayer under a $\mathcal{TP}$ operation]. Moreover, the sign-reversible L-MOE can be also excited in conventional $\mathcal{TP}$-symmetric AFMs under a vertical electric field, sharing the similar physics  to the novel dc case of layer Hall effect~\cite{AY-Gao2021}.}
	\label{fig:map}
\end{figure}

Magneto-optical Kerr~\cite{Kerr1877} and Faraday~\cite{Faraday1846} effects (MOKE, MOFE), referring to the change of polarization plane of light reflecting from and transmitting through a magnetic material, respectively (see Figure~\ref{fig:map}), offer prominent benefits for basic science and technology applications~\cite{Reim1990,Ebert1996,Antonov2004,Kuch2015}.
As a modern noncontact spectroscopic technique, magneto-optical effects (MOE) manifest itself as a powerful  tool for measuring magnetic order in two-dimensional (2D) materials, overcoming inevitable difficulties using standard techniques from superconducting quantum interference device and neutron scattering, as successfully performed in CrI$_3$~\cite{B-Huang2017}, Cr$_2$Ge$_2$Te$_6$~\cite{C-Gong2017}, and Fe$_3$GeTe$_2$~\cite{YJ-Deng2018}. Besides, MOE have also matured into a sophisticated probe for exploring band topology~\cite{L-Wu2016,Okada2016,Shuvaev2016,Dziom2017} and valley index~\cite{Lee2016,Lee2017}.

 The common MOE arises from the combined effects of band exchange splitting (induced by an finite net magnetization) and spin-orbit coupling (SOC)~\cite{Reim1990,Ebert1996,Antonov2004}. Antiferromagnets (AFMs), appear to pose a challenge for hosting MOE due to the absence of a net magnetization and usually suffer from the limitation from usual Kramers double band degeneracy. However, this empirical knowledge was questioned by several recent theoretical and experimental progresses in certain noncollinear AFMs. Feng \textit{et al.}~\cite{WX-Feng2015} first revealed an unexpected MOE in coplanar noncollinear AFMs Mn$_3X$ ($X$ = Rh, Ir, Pt) with special vector-spin chirality, which is also the cases for other kinds of noncollinear AFMs Mn$_3Z$N ($Z$ = Ga, Zn, Ag, Ni)~\cite{XD-Zhou2019a} and Mn$_3Y$ ($Y$ = Ge, Ga, Sn)~\cite{Wimmer2019}. The MOE in noncollinear AFMs was soon experimentally confirmed in Mn$_{3}$Sn~\cite{Higo2018,Balk2019} and Mn$_{3}$Ge~\cite{MX-Wu2020}. Subsequently, Feng \textit{et al.}~\cite{WX-Feng2020} further proposed a novel (quantum) topological MOE in noncoplanar AFMs, which stems from a finite scalar spin chirality and emerges even without SOC and band exchange splitting simultaneously, thus reshaping our understandings of the MOE. Very recently, the topological MOE was observed experimentally in skyrmion lattice~\cite{Kato2023}.

These advances indicate that the MOE is essentially controlled by magnetic symmetry~\cite{Sivadas2016}, with its symmetry rules traceable to the ac Hall pseudovector $\bm{\sigma}(\omega) = [\sigma_{yz}(\omega), \sigma_{zx}(\omega), \sigma_{xy}(\omega)]$~\cite{XD-Zhou2021}. The conventional collinear AFMs usually host $\mathcal{TP}$ or $\mathcal{T}\tau$  symmetry ($\mathcal{T}$ is the time-reversal; $\mathcal{P}$ is spatial inversion, and $\tau$ is lattice translation). The tensor $\bm{\sigma}$ is invariant under lattice translation. While with respect to $\mathcal{T}$ ($\mathcal{P}$) symmetry, the $\bm{\sigma}$ is odd (even) function. Thus, the combined $\mathcal{TP}$ or $\mathcal{T}\tau$ can force the $\bm{\sigma}$ to be zero and lead to vanishing MOE. For a 2D system ($\sigma_{yz} = 0$, $\sigma_{zx}  = 0$), the MOE is also suppressed by some $\mathcal{O}$ symmetries [e.g., $\mathcal{O}$ may be $\mathcal{M}_x$, $\mathcal{M}_y$, $C_{2x}$, and $C_{2y}$] as well as limited by some combined $\mathcal{TS}$ symmetries [e.g., $\mathcal{S}$ may be $\mathcal{M}_z$, $C_{2z}$, $C_{3z}$, $C_{4z}$, $C_{6z}$)] due to the odd property of $\sigma
_{xy}$ under these symmetries~\cite{Sivadas2016}. Hence, by breaking above symmetries, MOE may also emerge in more commonly available collinear AFMs~\cite{Sivadas2016,FR-Fan2017,K-Yang2020,Samanta2020,XD-Zhou2021,Mazin2021,N-Wang2022,Yananose2022,N-Ding2023}. From the nanoscience application perspective, it is much desirable to enrich the MOE family in 2D collinear AFMs because of the emergent valley-indexes~\cite{D-Xiao2007,D-Xiao2012,XD-Xu2014,WY-Tong2016,Mak2018,HY-Ma2021,XD-Zhou2021b,L-Liang2023}, layer physics~\cite{AY-Gao2021,ZM-Yu2020,Y-Ren2022,R-Chen2022,XY-Li2023,T-Zhang2023,YY-Feng2023,R-Peng2023}, and multiferroic order~\cite{CX-Huang2018,ML-Xu2020,JT-Zhang2020,XG-Liu2020,JY-Ji2023,P-Tang2023,YZ-Wu2023}. Importantly, it is highly desirable to realize the long-sought spin-splitting in 2D AFMs, which has triggered extensive attention recently benefiting from the versatile development of altermagnetism~\cite{Smejkal2022b,Smejkal2022a,XD-Zhou2024}.

In this work, we propose a layer control of MOE (L-MOE) in multiferroic bilayer Nb$_3$I$_8$ and topological bilayer MnBi$_2$Te$_4$.  The L-MOE is switched on and off by a layer stacking effect, depending on whether the $\mathcal{TP}$ symmetry is broken [Figure~\ref{fig:map}(c)-(d)] or not [Figure~\ref{fig:map}(a)-(b)]. The presence of L-MOE originates from the combined spin-valley-splitting in AFMs [Figure~\ref{fig:map}(d)], which enables the nonzero contributions from both the spin-conserving ($\uparrow\uparrow$ + $\downarrow\downarrow$) and spin-flip ($\uparrow\downarrow$) interband optical transitions. The spin-flip process is usually prohibited by the optical selection rule while it can be aroused in spin-splitting systems when considering the SOC effect~\cite{Weaver1979,Marusak1980,Khazan1993,Sexton1988,Hoerstel1998,HB-Zhang2011,Ezawa2012,Gibertini2014,Brotons2020,XD-Zhou2024}. It can be served as a unique hallmark for spin-valley splitting AFMs, a phenomenon absent in conventional $\mathcal{TP}$-symmetric AFMs due to the time-reversal symmetric interband transitions between states of opposing spins [$\uparrow\uparrow$ = - $\downarrow\downarrow$, Figure~\ref{fig:map}(b)]. The typical character of L-MOE is that its sign is reversible by reversing the  ferroelectric polarization or N{\'e}el vector directions [Figure~\ref{fig:map}(c)-(f)]. Moreover, we  reveal that the sign-reversible L-MOE can be also excited in the $\mathcal{TP}$-symmetric AFMs by an electric field effect, presenting the similar physics to the novel dc case of layer Hall effect~\cite{AY-Gao2021}. Interestingly, this sign-reversible L-MOE can be quantized in insulating topological bilayer antiferromagnets in the low-frequency limit. Our findings integrate the properties of emergent layer, spin, valley, ferroelectricity, and topology into the MOE and suggest a promising platform for spintronic, valleytronic, multiferroic, and magneto-optical applications.

\begin{figure}
\centering
	\includegraphics[width=0.7\columnwidth]{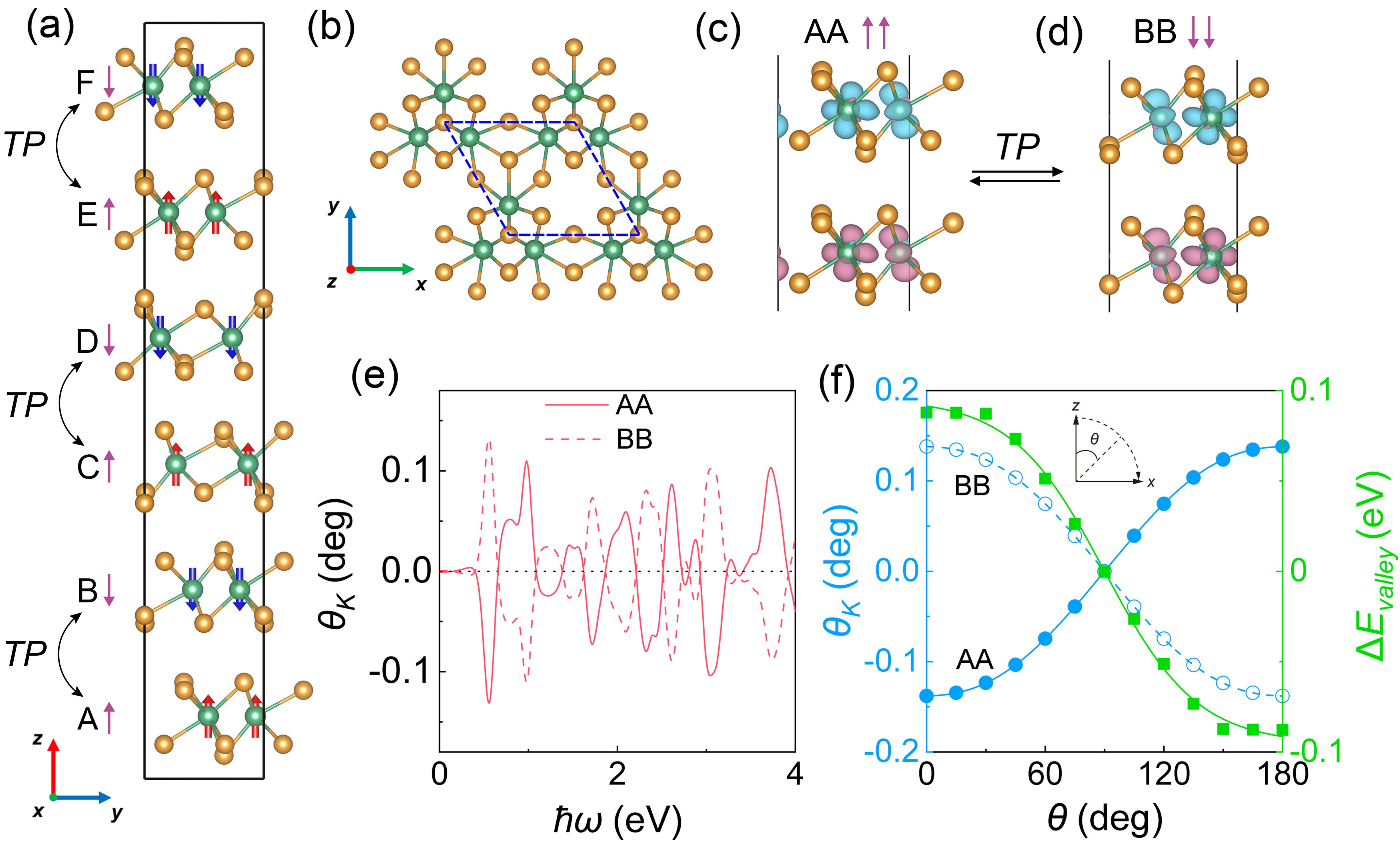}
	\caption{Sign-reversible L-MOE in multiferroic bilayer AFMs. (a) Side view of A-type antiferromagnetic bulk Nb$_3$I$_8$, labeled A-F from bottom to top layer, and (b) top view of monolayer Nb$_3$I$_8$. The red and blue arrows in (a) denote the magnetization direction. The A(C, E) and B(D, F) layers in the A-type antiferromagnetic bulk crystal are related to each other by a $\mathcal{TP}$ symmetry. The blue dash lines in (b) denote the primitive cell. (c-d) Side views of magnetic density isosurfaces for bilayer AA and BB stacked Nb$_3$I$_8$, which are related to each other by a $\mathcal{TP}$ symmetry and have an opposite ferroelectric polarization directions. The purple arrows in (a) and (c-d) indicate the polarization direction. The green and yellow balls in (a-d) represent Nb and I atoms, respectively. (e) L-MOE of bilayer Nb$_3$I$_8$ in AA and BB staking. (f) Angle evolution of maximum Kerr rotation angle (blue symbols and lines) and valley-splitting $\Delta E_{valley}$ (green symbols and lines) for AA and BB stacking. The $\Delta E_{valley}$ is defined as the energy difference between two $K$ valley at the conduction band edge [see Figure 4(g) and (l)].}
	\label{fig:crystal}
\end{figure}

The layered transition metal halides Nb$_3$\textit{X}$_8$ (\textit{X} = Cl, Br, I) have attracted increasing interest due to its exotic layered magnetic kagome lattice, topological flat bands, and transport properties~\cite{Simon1966,Magonov1993,Kim2018,Oh2020,JK-Jiang2017,R-Peng2020,Conte2020,Cantele2022,JY-Duan2024,L-Feng2023,J-Hong2023,Regmi2022,Regmi2023,Y-Zhang2023,SY-Gao2023}. Recently, the monolayer and few layers Nb$_3$I$_8$ were cleaved mechanically from the bulk crystal~\cite{Kim2018,Oh2020}. There are six monolayers in the bulk Nb$_3$I$_8$ with a stacking order of A-B-C-D-E-F along the $c$ direction [see Figure~\ref{fig:crystal}(a)]. Each layer hosts a breathing kagome lattice constructed by magnetic Nb trimers which are coordinated by  the distorted nonmagnetic I octahedron environment, as shown in Figure~\ref{fig:crystal}(b). Such an arrangement breaks inversion symmetry from space group of $P3m1$ and exhibits a spontaneous valley polarization~\cite{R-Peng2020} as well as ferroelectric polarization along the out-of-plane direction~\cite{L-Feng2023,J-Hong2023}. Among these, the A, C, and E monolayers are  interconnected through a lattice translation operation. They have ferroelectric polarization along the $z$ direction. The B (D, F) and A (C, E) layers are related to each other by an inversion symmetry (without considering magnetism). So, the ferroelectric polarization direction of B, D, and F monolayers is along the -$z$ direction, which are revealed by recent theoretical and experimental reports~\cite{L-Feng2023,J-Hong2023}. While the magnetism of Nb$_3$I$_8$ has yet to be measured experimentally, several preceding theoretical works have forecasted its layer-dependent magnetic properties~\cite{JK-Jiang2017,R-Peng2020,Conte2020,Cantele2022,L-Feng2023,JY-Duan2024}, e.g., ferromagnetism for monolayer and antiferromagnetism for bilayer. By comparing the total energies among the ferromagnetic, nonmagnetic, and different antiferromagnetic states [see Figures~S1-S2 and Tables~S1-S2], we find that the bulk and monolayer Nb$_3$I$_8$ prefer to a A-type  antiferromagnetic ground state [see Figure~\ref{fig:crystal}(a)] and ferromagnetic ground state, respectively, which is in accordance with the prior studies~\cite{JK-Jiang2017,R-Peng2020,Conte2020,Cantele2022,L-Feng2023,JY-Duan2024}. The optimized lattice constants of monolayer, bulk crystal, and bilayers are shown in Table~S1-S3, which are in reasonable agreement with the experimental results~\cite{Simon1966,Magonov1993}.

The key findings of L-MOE are demonstrated in the bilayer multiferroic Nb$_3$I$_8$. The previous studies~\cite{L-Feng2023,J-Hong2023} suggest that the two layers with same (opposite) polarization directions are the ferroelectric (antiferroelectric) states. For example, the AA (CC, EE) bilayer hosts the ferroelectric polarization along the $z$ direction, while the BB (DD, FF) being along the $-z$ direction [see Figure~\ref{fig:crystal}(c)-(d)], respectively.  Correspondingly, the AB (BC, CD, DE, and EF) layer stacking is the antiferroelectric state. By layer sliding and twisting, one can obtain more abundant multiferroic states in bilayer Nb$_3$I$_8$. Here, we consider seven simplest stacking configurations such as AA(BB), AB, AC, AD, AE, and AF, which are sufficient to reveal the L-MOE in the following discussion. The magnetic ground states of AD and AE prefer to interlayer ferromagnetic coupling, while AA (BB), AB, AC, and AF feature an interlayer A-type antiferromagnetic coupling, showing a strong stacking-dependent magnetism which is also reported in bilayer CrI$_3$~\cite{Sivadas2018} and MnBi$_2$Te$_4$~\cite{Y-Ren2022}. Next, we will focus on bilayer antiferromagnetic AA (BB), AB, AC, and AF stacking to reveal the sign-reversible L-MOE.

First, the basic property of sign-reversible L-MOE is demonstrated directly based on the magnetization density and magnetic group analyses. Figure~\ref{fig:crystal}(c) shows the magnetization density of bilayer AA stacking, the usual $\mathcal{TP}$, $\mathcal{TS}$, and $\mathcal{O}$ symmetries that suppress the MOE are broken apparently. In this case, the $\sigma_{xy}$ or L-MOE is nonzero protected by the magnetic point group of $3m^{\prime}1$. Interestingly, the ferroelectric counterpart of BB stacking has a degenerate energy and same magnetic point group as the AA stacking, while their magnetic density isosurfaces are connected by a $\mathcal{TP}$ transformation [see Figure~\ref{fig:crystal}(c) and (d)]. The MOE is odd and even with respect to $\mathcal{T}$ and $\mathcal{P}$ symmetries, respectively. Therefore, the two opposite antiferromagnetic ferroelectric states (AA and BB) must have a reverse sign of L-MOE. This sign-reversible principle can also be extended into all other paired ferroelectric states of bilayer Nb$_3$I$_8$ with opposite polarization directions. The magnetic group analyses are further confirmed by the first-principles results. The Figure~\ref{fig:crystal}(e) presents a long-sought MOE in AA- and BB-stacked antiferromagnetic Nb$_3$I$_8$ with the N{\'e}el vector beling along the $z$ direction. The largest Kerr rotation angle of about 0.13$^\circ$ is comparable to the typical magnitude of $0^{\circ}\sim2^\circ$ in some usual layered magnets, such as CrI$_3$~\cite{Gudelli2019}, Cr$_2$Ge$_2$Te$_6$~\cite{YM-Fang2018}, Fe$_n$GeTe$_2$ ($n$ = 3, 4, 5)~\cite{XX-Yang2021a}, CrTe$_2$~\cite{XX-Yang2021}, and CrSBr~\cite{XX-Yang2022}. As expected from symmetry analyses, the AA and BB stacking with opposite ferroelectric polarization directions exhibit the same magnitude but opposite sign of L-MOE. 

\begin{figure}
\centering
	\includegraphics[width=0.65\columnwidth]{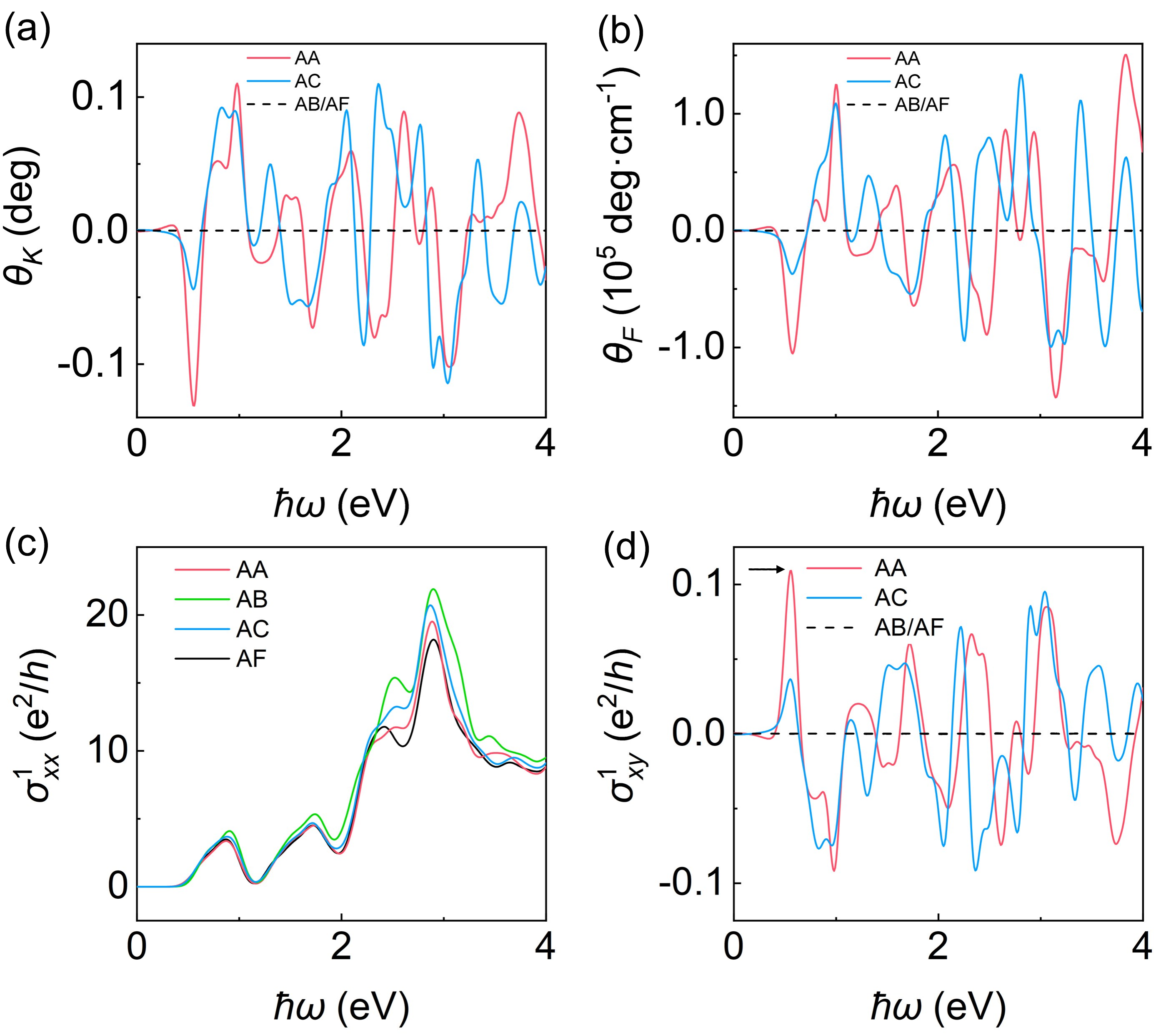}
	\caption{Layer-stacking dependent L-MOE in bilayer AFMs. (a-b) MOKE and MOFE, (c-d) real parts of diagonal element $\sigma_{xx}$ and off-diagonal element $\sigma_{xy}$, for bilayer antiferromagnetic Nb$_3$I$_8$ with different stacking. The arrow in (d) marks the maximum of $\sigma_{xy}^{1}$.}
	\label{fig:layer}
\end{figure}

Besides the induced sign-reversible response, the layer stacking also plays a significant role in the presence and absence of L-MOE in bilayer Nb$_3$I$_8$. The L-MOE of AB and AF stacking are vanishing because the systems possess a $\mathcal{TP}$ symmetry from a magnetic point group of $\bar{3}^{\prime}1m^{\prime}$. While for AC stacking, the L-MOE is allowed due to the same magnetic point group $3m^{\prime}1$ as the AA stacking. The first-principles calculations indeed reproduce the results of symmetry analyses. Figure~\ref{fig:layer}(a-b) shows the magneto-optical Kerr and Faraday rotation angles $\theta_{K, F}$ for different layer stacking. Apparently, the $\theta_{K, F}$ for AB and AF stacking are equal to zero in the whole frequency range, while it is nonvanishing for AA and AC stacking. One can also observe that the magnitudes of L-MOE are substantially affected by layer stacking. This novel phenomenon is essentially controlled by the off-diagonal element $\sigma_{xy}$ rather than diagonal element $\sigma_{xx}$ of optical conductivity. The former (latter)  depends significantly on (is robust against) layer stacking, as shown in Figure~\ref{fig:layer}(c-d). 

\begin{figure}[t]
\centering
	\includegraphics[width=\columnwidth]{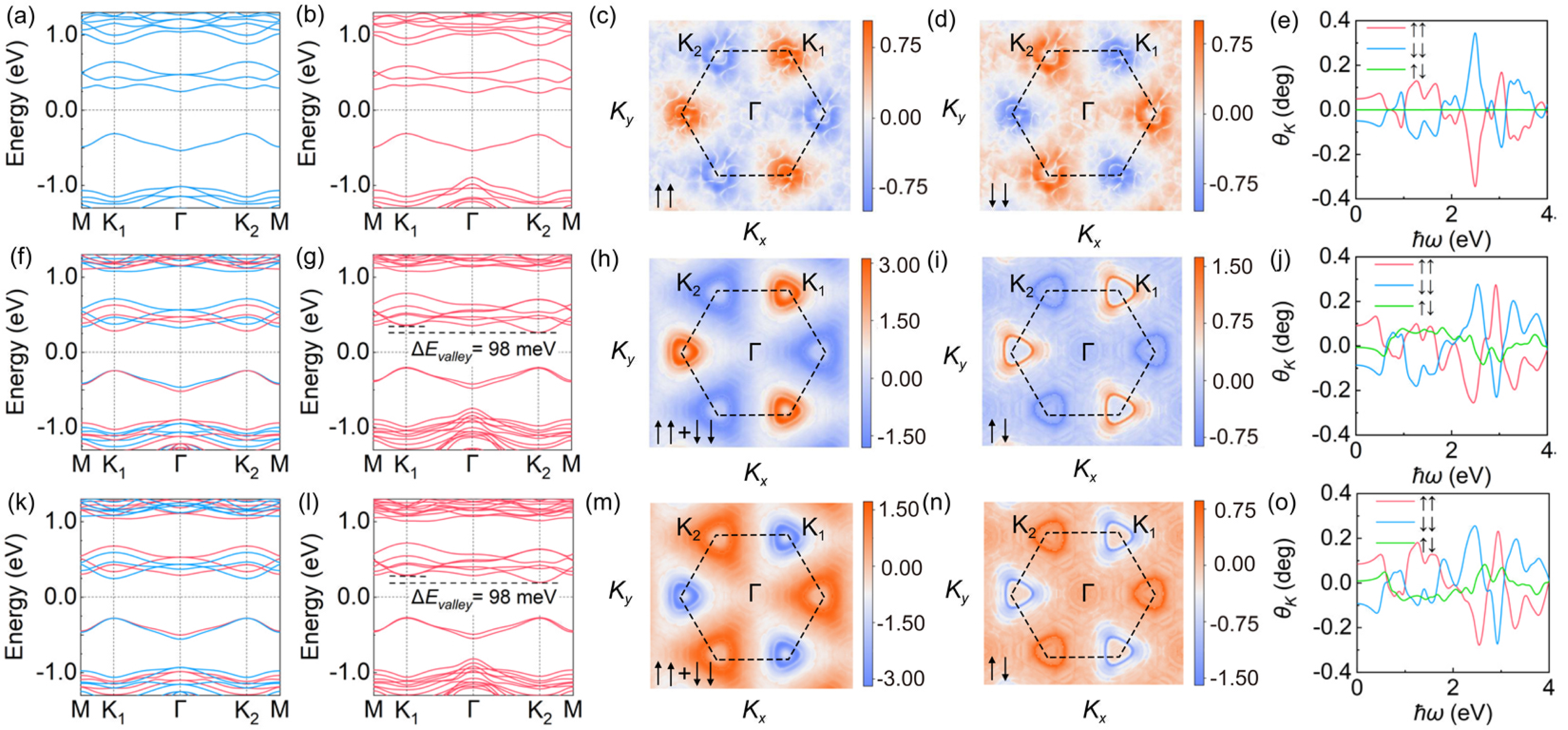}
	\caption{Underlying physical origin of sign-reversible L-MOE. (a, b)/(f, g) Spin-polarized and relativistic band structures of bilayer Nb$_3$I$_8$ with AB/AA stacking, respectively. (c, d) Momentum-resolved $\sigma_{xy}^{1}$ for AB stacking with the interband optical transitions between the bands from spin-up to spin-up ($\uparrow\uparrow$), and spin-down to spin-down ($\downarrow\downarrow$), respectively. (h, i) Similar to (c, d) but from spin-conserving ($\uparrow\uparrow$ + $\downarrow\downarrow$) and spin-flip ($\uparrow\downarrow$) interband transitions. (e)/(j) Spin-resolved L-MOE with AB/AA stacking contributed from the interband transitions of $\uparrow\uparrow$, $\downarrow\downarrow$, and spin flip $\uparrow\downarrow$, respectively. (k-o) Similar to (f-j) but with BB stacking. In (c, d), (h, i) and (m, n), the $\sigma_{xy}^{1}$ (in units of $e^2/h$) is calculated at $\hbar\omega$ = 0.56 eV and is marked by a black arrow in Figure~\ref{fig:layer}(d). The reduced coordinates for the Berry curvature slice and high-symmetry points are shown in Figure~S9.}
	\label{fig:spin_valley} 
\end{figure}

Taking the AA, BB, and AB stackings as typical examples, we now elucidate the underlining physics of the presence/absence of L-MOE and their sign-reversible response at the microscopic level based on the electronic structure calculations. Figure~\ref{fig:spin_valley}(a) and (b) show the spin-polarized and relativistic band structures of AB stacking with the N{\'e}el vector being along the $z$ direction, respectively. Without SOC, one clearly observes that the AB stacking obeys Kramers spin degeneracy with a pair of degenerate valleys at K$_1$ and K$_2$ for both conduction and valence band edges. After considering the SOC effect, the double band degeneracy still exists due to the $\mathcal{TP}$ symmetry, while the valley degeneracy is broken due to the novel coupled spin-valley physics~\cite{X-Li2013}. By decomposing the interband optical transitions between the bands of the same spins (namely, spin-conserved parts $\uparrow\uparrow$ or $\downarrow\downarrow$) and the opposite spins (namely, spin-flip process $\uparrow\downarrow$), one can see that the momentum-resolved $\sigma_{xy}^{1}$ for spin-up to spin up ($\uparrow\uparrow$) and spin-down to spin-down ($\downarrow\downarrow$) interband transitions present a reverse distributions around two K valleys [see Figure~\ref{fig:spin_valley}(c) and (d)]. Thus, for this conventional AFM, the magneto-optical spectra from spin-conserved parts are cancelled out from each other, as shown in Figure~\ref{fig:spin_valley}(e), accompanying with a negligible
contribution from the spin-flip process.

Nevertheless, a novel L-MOE is generated in AA stacking due to the lifted spin  together with valley degeneracy, as revealed in the following discussion. Figure~\ref{fig:spin_valley}(f) and (g) show the spin-polarized and relativistic band structures of AA stacking with the N{\'e}el vector being along the $z$ direction, respectively. It is obvious that the Kramers spin degeneracy is lifted in the entire Brillouin zone even in this fully compensated collinear AFM, in contrast to the spin-momentum locking property in altermagnet~\cite{Smejkal2022b,Smejkal2022a,XD-Zhou2024}. Therefore, the multiferroic bilayer Nb$_3$I$_8$ can also host the spin-polarized property of ferromagnets coupled with the extraordinary advantages of antiferromagnetic spin order such as ultrafast spin dynamics and without parasitic stray fields. The SOC effect further breaks the degeneracy of two K valleys. The valley splitting $\Delta E_{valley}$ can reach up to 98 meV at the conduction band edge, which is on the order of usual ferrovalley materials, such as monolayer Nb$_3$I$_8$~\cite{R-Peng2020}, VSe$_2$ or VS$_2$~\cite{WY-Tong2016,XG-Liu2020}, VSi$_2$N$_4$ or VSi$_2$P$_4$~\cite{QR-Cui2021,XD-Zhou2021b,XY-Feng2021,L-Liang2023,FY-Zhan2023,SD-Guo2023}. In contrast to $\mathcal{TP}$-symmetric antiferromagnetic materials, Figure~\ref{fig:spin_valley}(h) shows that the spin-conserving interband transitions ($\uparrow\uparrow$+$\downarrow\downarrow$) present an asymmetric distribution near two K valleys, which means the $\uparrow\uparrow$ component can not cancel out the $\downarrow\downarrow$ contribution. Notably, a finite spin-flip contribution emerges and also exhibits an unbalanced distributions around two K valleys [see Figure~\ref{fig:spin_valley}(i)], which is usually negligible small in most materials because the SOC energy scale is usually smaller than the exchange splitting energy scale~\cite{Weaver1979,Marusak1980,Khazan1993,Sexton1988,Hoerstel1998,HB-Zhang2011,Ezawa2012,Gibertini2014,Brotons2020,XD-Zhou2024}. The nonvanishing contributions both from spin-conserving and spin-flip processes, induced by the strong SOC effect and special spin-splitting, ultimately give rise to a measurable L-MOE, as shown in Figure~\ref{fig:spin_valley}(j). One can see that the novel spin-flip process can be served as a unique marker for spin-valley splitting AFMs, which is absent for conventional $\mathcal{TP}$-symmetric AFMs. The sign-reversible magneto-optical response for AA and BB stacking can be also elucidated based on the spin-valley splitting physics. Figure~\ref{fig:spin_valley}(k)-(l) shows the spin-polarized and relativistic band structures for BB stacking. Apparently, they can be regarded as the time-reversal-like counterpart of the case in AA stacking. The $\sigma_{xy}$ and MOE are all odd under time-reversal symmetry. Therefore, the signs for all of them are reversed when the ferroelectric polarization direction is changed, as shown in Figure~\ref{fig:spin_valley}(m)-(o).

We also find that the L-MOE can be effectively tuned by changing magnetization direction [see Figure~\ref{fig:crystal}(f)], which is strongly correlated to the valley polarization property since they all originate from the SOC effect. With the N{\'e}el vector $\bm{n}$ rotating from the $z$-axis to the -$z$-axis gradually, the L-MOE and valley-splitting $\Delta E_{valley}$ all display a period of 2$\pi$ in $\theta$. This can be understood from that the $z$ component of spins at $\pi \pm \theta$ is the time-reversed counterpart of the one at $\theta$ and the L-MOE/$\Delta E_{valley}$ is odd with respect to time-reversal symmetry $\mathcal{T}$. Especially, when the N{\'e}el vector is along the $x$ direction, although the spin-splitting exists, while the valley-splitting is vanishing and L-MOE is thus prohibited [see Figure~S6]. The underlining physics is that the system has a $\mathcal{O} = \mathcal{M}_x$ symmetry from magnetic point group of $m$ when N{\'e}el vector $\bm{n} || x$. Because $\sigma_{xy}$ is odd under $\mathcal{M}_x$ symmetry, the L-MOE is expected to be zero. Thus, the predicted L-MOE is closely related to both the spin- and valley splitting, and it is characterized by the sign-reversible response controlled by ferroelectric polarization together with N{\'e}el vector directions. 

Beyond the layer-stacking method, the L-MOE can be also facilitated even for $\mathcal{TP}$-symmetric AFMs (such as for bilayer Nb$_3$I$_8$ with AB stacking) via applying a static electric field, which can break the $\mathcal{TP}$ symmetry and induce the spin-splitting~\cite{Sivadas2016}. The combined spin- and valley-splitting leads to the gate-controllable L-MOE [seeing Figures~S10 and S16(a)]. Moreover, this field-driven L-MOE can be extended into a quantized version in antiferromagnetic Chern insulators, as done in bilayer MnBi$_2$Te$_4$ in the following discussion. It is known that the even-layer MnBi$_2$Te$_4$~\cite{JH-Li2019,Otrokov20191,DQ-Zhang2019,C-Liu2020,YJ-Deng2020,J-Ge2020} is a typical antiferromagnetic axion insulator with $\mathcal{TP}$-symmetric Kramers spin degeneracy, which provides a ideal platform for investigating the transport phenomena under the axion field \textbf{E}$\cdot$\textbf{B} effect. \textit{Gao et al.}~\cite{AY-Gao2021} observed experimentally a fascinating type of Hall effect, coined layer Hall effect, in which the compensated layer-locked Berry curvature's contribution from top and bottom layer is broken and reversed by the axion field of \textbf{E}$\cdot$\textbf{B}. Our calculated layer Hall conductivity can reach the same order  of the experimental report (\textit{i.e.}, 0.5 $e^2/h$)~\cite{AY-Gao2021} at appropriate hole- or electron-doping, as shown in Figure~S14. By extending the layer Hall effect into the ac case, it is natural to expect the sign-reversible L-MOE controlled by the electric field and N{\'e}el vector. 

\begin{figure}[t]
	\includegraphics[width=0.8\columnwidth]{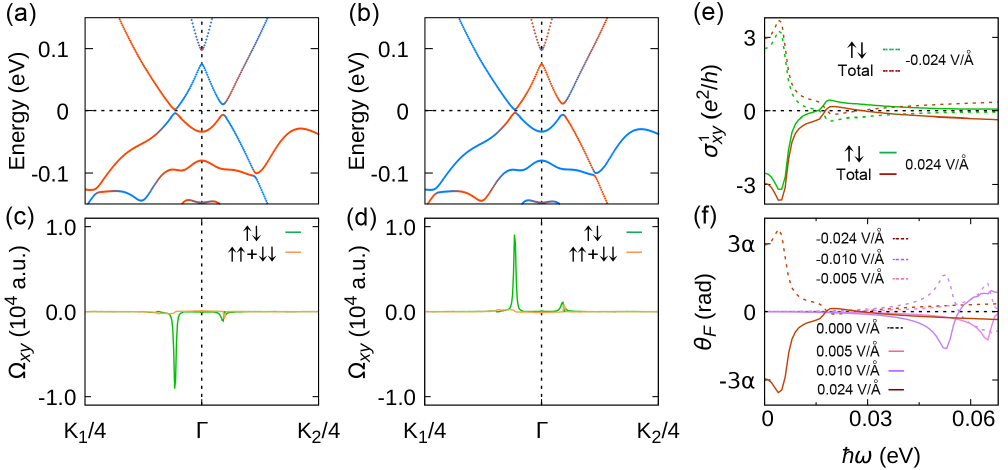}
	\caption{Sign-reversible quantum L-MOE in topological bilayer AFMs. (a, b) Band structure with SOC for bilayer MnBi$_2$Te$_4$ at $E = \pm 0.024$ V/\AA. The colors denote the bands of the spin projection along the $z$ direction. (c, d) Spin-resolved Berry curvature along symmetry lines at $E = \pm 0.024$ V/\AA \ from spin flip transition $\uparrow\downarrow$, and spin-conserved transition ($\uparrow\uparrow$+$\downarrow\downarrow$), respectively. (e) Spin-resolved optical Hall conductivity $\sigma_{xy}^1$ at $E = \pm 0.024$ V/\AA \ from spin flip transition $\uparrow\downarrow$, and the total contribution ($\uparrow\downarrow$ + $\uparrow\uparrow$ + $\downarrow\downarrow$), respectively. (f) L-MOE and its quantization of bilayer MnBi$_2$Te$_4$ at varying electric field. }
	\label{fig:QMO} 
\end{figure}

Interestingly, a novel topological phase transition is induced in bilayer MnBi$_2$Te$_4$ with the an increase of the electric field or SOC strength, as shown in Figure~S13. Specially, a Chern insulator phase emerges at a critical field of $E \approx 0.023-0.027$ V/\AA~\cite{SQ-Du2020} even without net magnetization. Because the role of SOC strength is closely related to the known Berry curvature (behaving like a fictitious magnetic field), the emergence of the topological phase transition may be traceable to the combined control by electric and magnetic field, which in fact has been first realized experimentally recently by Cai \textit{et al.} in spin-canting MnBi$_2$Te$_4$~\cite{JQ-Cai2022}. The Chern insulator phase in bilayer MnBi$_2$Te$_4$ is characterized by a large quantized optical Hall conductivity in the low-frequency limit $\sigma^{1}_{xy}= -3e^2/h$ [see Figure~\ref{fig:QMO}(e)]. As expected from the physics of layer Hall effect~\cite{AY-Gao2021}, the quantized $\sigma^{1}_{xy}$ is also reversed by changing the electric field direction, which arises from the fact that the system at $E =\pm$ 0.024 V/\AA~hosts time-reversal-like degenerate spin-polarized electronic bands and the resulting opposite Berry curvatures, as shown in Figure~\ref{fig:QMO}(a)-(d). We also find that the large quantized $\sigma^{1}_{xy}$ is mainly contributed by the Berry curvature along the $\Gamma \rightarrow K_1$ direction [see Figure~\ref{fig:QMO}(c)-(d)], which is more than 4 orders of magnitude larger than that along the $\Gamma \rightarrow K_2$ direction, showing a valley-like polarized feature. Such a valley property, located at the general paths rather than high-symmetry points, is also reported in few-layer transition metal dichalcogenides~\cite{ZF-Wu2016}. Much interestingly, the Berry curvature and the $\sigma^{1}_{xy}$ mainly stem from the contribution from spin-flip transition [see Figure~\ref{fig:QMO}(c)-(e)] because the strong SOC-induced avoided crossings are separated between two bands with opposite spin states~\cite{Gibertini2014} [see Figure~\ref{fig:QMO}(a)-(b)].  Overall, the field effect shares the similar underlining physics as the above layer stacking effect, that is the spin-valley splitting in $\mathcal{TP}$-broken bilayer AFMs.

This spin-valley-polarized properties together with the topological properties ultimately give rise to the sign-reversible L-MOE in bilayer MnBi$_2$Te$_4$ [see Figure~\ref{fig:QMO}(f)]. In addition, the sign-reversible L-MOE is quantized in the low-frequency limit, $i.e.$, $\theta_{K}=  -\textrm{tan}^{-1}[c/(2\pi\sigma^{1}_{xy})] \simeq \pm \pi/2$ [see Figure~S16(b)] and $\theta_{F} = \textrm{tan}^{-1}(2\pi\sigma^{1}_{xy}/c) \simeq C\alpha$ [see Figure~\ref{fig:QMO}(f)] ($c$ is the speed of light, $C$ = $\pm$3 is the Chern number) due to the consideration from an axion term $(\Theta\alpha/4\pi^{2})$\textbf{E}$\cdot$\textbf{B} (here $\Theta$ is magnetoelectric polarizability and $\alpha =  \text{e}^{2}/\hbar c$ is fine structure constant) into the usual Maxwell’s equations~\cite{XL-Qi2008,Maciejko2010,Tse2010,Tse2011}. Thus, our findings are a crucial step for detecting experimetally this novel spin-valley polarized physics and the hidden topology simultaneously in such a large family of bilayer collinear AFMs by utilizing the noncontact modern magneto-optical technique~\cite{B-Huang2017,C-Gong2017,L-Wu2016,Okada2016,Shuvaev2016,Dziom2017}.

In this study, using symmetry analyses and first-principles calculations, we proposed a novel class of magneto-optical interactions, namely L-MOE, which is rooted in the nontrivial combination of spin- and valley-splitting in bilayer AFMs. The L-MOE can be switched on/off by a layer-stacking effect or an electric field stimulus, originating from the uncompensated/compensated spin-conserving and spin-flip interband transitions. The key characteristic of L-MOE is the sign-reversible responses controlled by ferroelectric polarization, the N{\'e}el vector, or the electric field direction. Interestingly, the sign-reversible L-MOE can be quantized in topologically insulating AFMs. These findings promote our understanding of MOE in a large family of bilayer AFMs, and may have a potential impact in magneto-optical devices based on the novel layer, spin, and valley degrees of freedom.

\begin{suppinfo}	
		The Supporting Information contains: (a) the details of methods, (b) supplementary tables, figures, and discussions on other calculation results including different magnetic configurations, lattice constants, the dependence of Hubbard parameter $U$, band structures with different layer stacking and magnetization, transition between ferroelectric state and antiferroelectric state, (spin-resolved) optical conductivity with different layer stacking in bilayer Nb$_3$I$_8$ as well as the reduced coordinates for the Berry curvature slice, band structures and magneto-optical effects under electric fields, topological phase transition under different SOC and electric field strength, layer Hall conductivity and optical Hall conductivity in bilayer MnBi$_2$Te$_4$.		
\end{suppinfo}	

	\begin{acknowledgement}
The authors thank Wanxiang Feng, Li Feng, Run-Wu Zhang, Gui-Bin Liu, Ping Yang, and Libor {\v S}mejkal for fruitful discussions. This work is supported by the National Natural Science Foundation of China (Grants Nos. 12304066, 12074154, 11722433, and 12174160), the Natural Science Foundation of Jiangsu Province (Grant No. BK20230684), the Natural Science Fund for Colleges and Universities in Jiangsu Province (Grant No. 23KJB140008), Six Talent Peaks Project and 333 High-level Talents Project of Jiangsu Province.
	\end{acknowledgement}

\bibliography{references}

\end{document}